\documentclass[useAMS,usenatbib]{emulateapj} 
\usepackage{epsfig}
\usepackage{lineno}
\usepackage{epsfig}  
\usepackage{amsmath} 
\usepackage{amssymb}
\usepackage{natbib} 
\usepackage{graphicx} 
\usepackage{endnotes}
\usepackage{color}

\shorttitle{ Milli-second phenomena in  4U 1728-34} \shortauthors{ Chauhan et al.}

\begin{document}

\title{ {\it AstroSat}/LAXPC detection of milli-second phenomena in 4U 1728-34} 
\author{Jai Verdhan Chauhan$^1$, J S Yadav$^1$, Ranjeev Misra$^2$, P C Agrawal$^3$, H M Antia$^1$, Mayukh Pahari$^2$, Navin Sridhar$^4$, Dhiraj Dedhia$^1$, Tilak Katoch$^1$, P. Madhwani$^1$, R K Manchanda$^5$, B Paul$^6$, Parag Shah$^1$} \affil{$^1$ Tata Institute of Fundamental Research, Homi Bhabha Road, Mumbai, India
\texttt{jai.chauhan@tifr.res.in}} \affil{$^2$ Inter-University Centre for Astronomy and Astrophysics, Pune 411007, India} \affil{$^3$ {  {UM-DAE Center of Excellence for Basic Sciences, University of Mumbai, Kalina, Mumbai-400098, India}}} \affil{$^4$ {  {Indian Institute of Science Education and Research, Bhauri, Bhopal 462006, India}}} \affil{$^5$ University of Mumbai, Kalina, Mumbai-400098, India} \affil{$^6$ Dept. of Astronomy \& Astrophysics, Raman Research Institute,  Bengaluru-560080 India }

\begin{abstract}

{  {The low mass X-ray binary 4U 1728-24  was observed
    with {\it AstroSat}/LAXPC on 8th March 2016. Data from a randomly
    chosen one orbit of over 3 ks was analyzed for detection of rapid
    intensity variations. We found that the source intensity was
    nearly steady but towards the end of the observation a typical
    Type-1 burst was detected.  Dynamical power spectrum 
      of the data {  in the 3-20 keV band,}
    reveals presence of a kHz Quasi-Periodic Oscillation (QPO) whose
    frequency drifted from $\sim 815$ Hz at the beginning of the
    observation to about 850 Hz just before the burst.  
    {  The QPO is also detected in the 10-20 keV band, which
was not obtainable by earlier RXTE observations of this source.} Even for such a
    short observation with a drifting QPO frequency, the time-lag
    between the 5--10 and 10--20 keV bands can be constrained to be
    less than 100 microseconds. The Type-1 burst that lasted for about
    20 secs  had a typical profile.   {blue}During the first four seconds
    dynamic power spectra reveal a burst oscillation whose frequency increased from 
 $\sim 361.5$ to $\sim 363.5$Hz}. This is consistent with the earlier results
    obtained with {\it RXTE}/PCA, showing the same spin frequency of
    the neutron star. The present results demonstrate the capability
    of LAXPC instrument for detecting millisecond variability even from
    short observations. After {\it RXTE} ceased operation, LAXPC
    on {\it AstroSat} is the only instrument at present with
    capability of detecting kHz QPOs and other kind of rapid
    variations from 3 keV to 20 keV and possibly at higher energies
    also.}

\end{abstract}


\section{Introduction}\label{intro}

One of the most { {important and lasting legacy}} of the {\it Rossi X-ray Timing Experiment} ({\it RXTE}) has been the discovery and characterization of milli-second phenomena in X-ray binaries \citep[e.g.][]{Van00,Van06,Rem06}. These include the detection of kilo-Hertz Quasi-periodic  Oscillations (QPO) and the coherent Burst Oscillation (BO) in the initial phase of  Type-1 (or thermo-nuclear) bursts. 

Since their discovery which occurred soon after the launch of {\it RXTE}, kHz QPOs have been the subject of extensive { {research and discussion.}} The high frequency of the variability, implies that the phenomena is linked to the behaviour of matter in the inner edge of the accretion disk close to the neutron star surface, and hence has the promise of revealing the behaviour of matter in the strong gravitational field limit. { {In several low mass X-ray binaries}} these QPOs have been observed in pairs and there have been several  detailed { { studies}} of their occurrence and the { {relationship}} between the pairs { {of frequencies}} as well as with that of other low frequency QPOs \citep[e.g.][]{Str03,Alt08,Men98,Bel07}. Attempts have been made to explain these relations with theoretically motivated models where one of the frequencies is identified { {as}} a Keplerian one and the other a result of complex interactions that may occur in such regions, such as beating { {of frequencies}} or resonances \citep[e.g.][]{Lam98,Ste99,Osh99}. Despite these endeavours, it is perhaps fair to say that there is no consensus on which of these dynamical models represents the correct physical picture, and { {the underlying physical phenomena of kHz QPO remains an open question. While these studies have focused on understanding the dynamical origin of the kHz QPOs, there have been relatively less attempts}} on identifying the radiative processes by which the phenomena is manifested. The QPOs are known to occur at particular spectral states of the system and understanding the radiative components and specifically the spectral parameters that vary to produce the QPO, would enhance our understanding of their origin. This can be done by studying the fractional r.m.s  and time lag as a function of photon energy \citep[e.g.][]{Ber96,Men01,Vau97,Kaa99,Bar13,Ave13,Pei15}. The increase of r.m.s with energy and soft lags can be explained in the framework of a Comptonization model  and such analysis 
can not only constrain the responsible radiative process but also { {provide estimates of the size}} and geometry of the region \citep{Lee01,Kum16}.

    It should be emphasized that { {a large fraction of
        the X-ray binaries}} are transient and the kHz QPOs
    { {are known to occur}} only during certain spectral
    states. Hence it is important to continuously monitor the X-ray
    sky for new X-ray binaries and to study the known ones to get new
    insights { {in}} to the phenomena. For example,
    oscillations observed on { {time-scales
        significantly smaller than mill-seconds, will revolutionize
        our understanding of the kHz QPOs.}} Thus, there is a critical
    need for instruments { {that have the capability of
        observing the high frequency variations in the post-{\it RXTE}
        era.  }

Unlike  kHz QPOs, the frequency of the coherent oscillation observed during the initial { {burst decay}} phase of a Type-1 burst, is  unambiguously related to the spin of the neutron star \citep[e.g.][]{Str97,Cha03}. During the initial period of a Type I burst, it is believed that uneven nuclear burning of material accreted on  the surface of the neutron star, is the cause of the observed coherent burst { {oscillations,}} although the exact mechanism is not clear \citep[e.g.][]{Str99,Mun04,Cha14}. { {Indeed, for low mass X-ray binaries, our knowledge of the spin period of the neutron star is solely derived from the burst oscillations}.} Thus, for new X-ray transients, it is important to have the capability to measure the BO and hence { {infer the spin period of the neutron star. The Type I bursts and the rapid oscillations}} are by themselves an interesting phenomenon, { {providing rich information about the nuclear burning process and other}} several not well understood features like why the oscillations persists for as long as 5-10 s. Broad band observations of these bursts and energy dependent time-lags will provide crucial information { {to probe these processes deeper.}}

With an effective area similar to that of {\it RXTE} at low $\sim 5$
keV and significantly larger at higher energies ($\sim 50$ keV), the
Large Area X-ray Proportional Counter (LAXPC)
\citep{Yad16,Ant16} on board the Indian X-ray mission, {\it
  AstroSat} \citep{Agr06,Sin14}, is expected to detect and discover
milli-second variability in X-ray binaries, leading to significant
enhancement { {of}} our understanding of the
phenomena. Moreover, the other instruments on board {\it AstroSat}
will provide wide band coverage from UV to hard X-rays. LAXPC data of
black hole systems GRS1915+105 and Cygnus X-1 have already
{ {demonstrated the capability} of LAXPC to study
  variability of high energy photons \citep{Yad16b,Mis16}.

4U 1728-34 (GX 354-0) is a well studied atoll type low mass X-ray
binary for which {\it RXTE} has detected kHz QPOs during several
occasions \citep[e.g.][]{Mig03,Str96,Muk12}.  The frequency of the
lower kHz QPO covers a wide range from 300 to 1100 Hz.  The source
exhibits frequent Type-1 bursts for which burst oscillations have been
detected at $\sim 363$ Hz \citep{Str97} and extensively studied
\citep{Str01,Mun01,Mun04,Zha16}. Here we report the first detection of
both kHz QPO and the burst oscillation for a short $\sim 3$ ksec
observation of 4U 1728-34 by {\it AstroSat}/LAXPC.  The kHz QPO 
is detected in energies $> 10$ keV which RXTE was not able to do \citep{Muk12}, thereby
convincingly demonstrating its superior 
capability to detect milli-second variability.

\section{Detection of kHz QPO}

\begin{figure}
\centering
\includegraphics[width=0.34\textwidth,angle=-90]{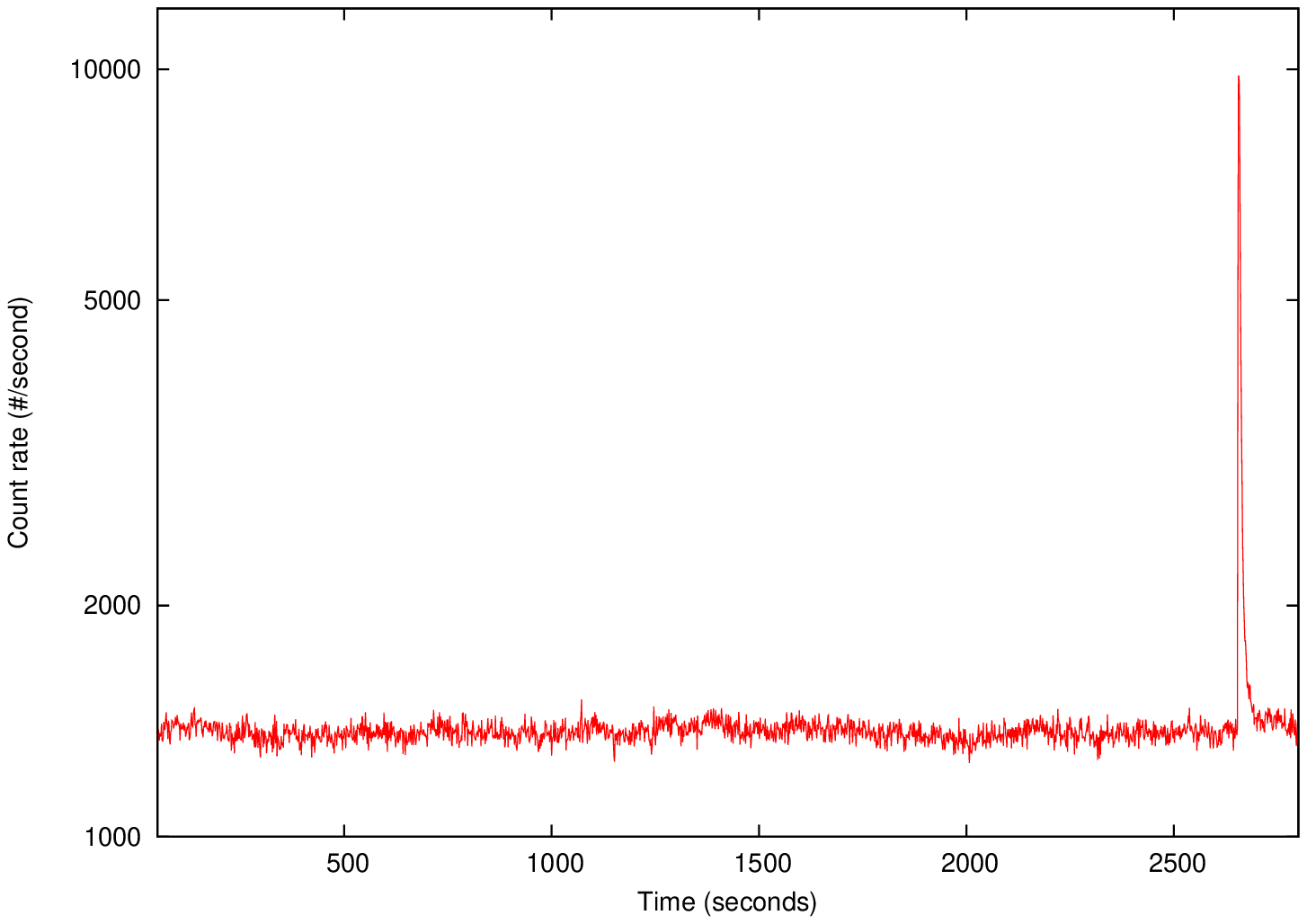}
\caption{Lightcurve of  4U 1728-34 in the energy
range 3--20 keV is shown where the count rate from all three LAXPC
detectors are combined. }
\label{lightcurve}
\end{figure}

\begin{figure}
\centering
\includegraphics[width=0.34\textwidth,angle=-90]{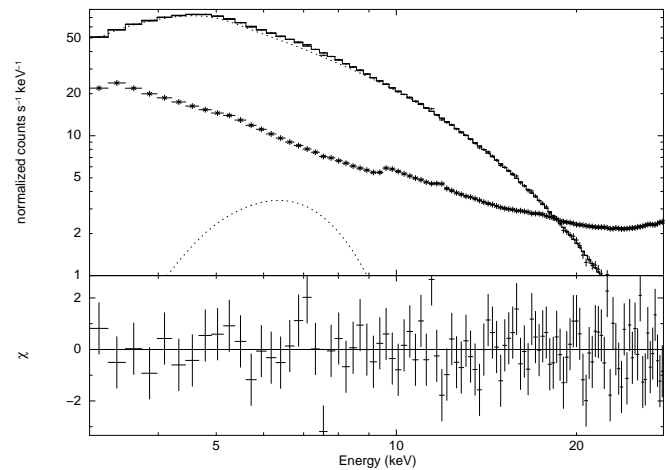}
\caption{ The photon spectrum from LAXPC 10 for the first $\sim 2500$ seconds.
The spectrum is modelled using a thermal Comptonization component and a broad
Iron line. The plot also shows the expected background spectrum}
\label{spec}
\end{figure}

\begin{figure}
\centering
\includegraphics[width=0.3\textwidth,angle=-90]{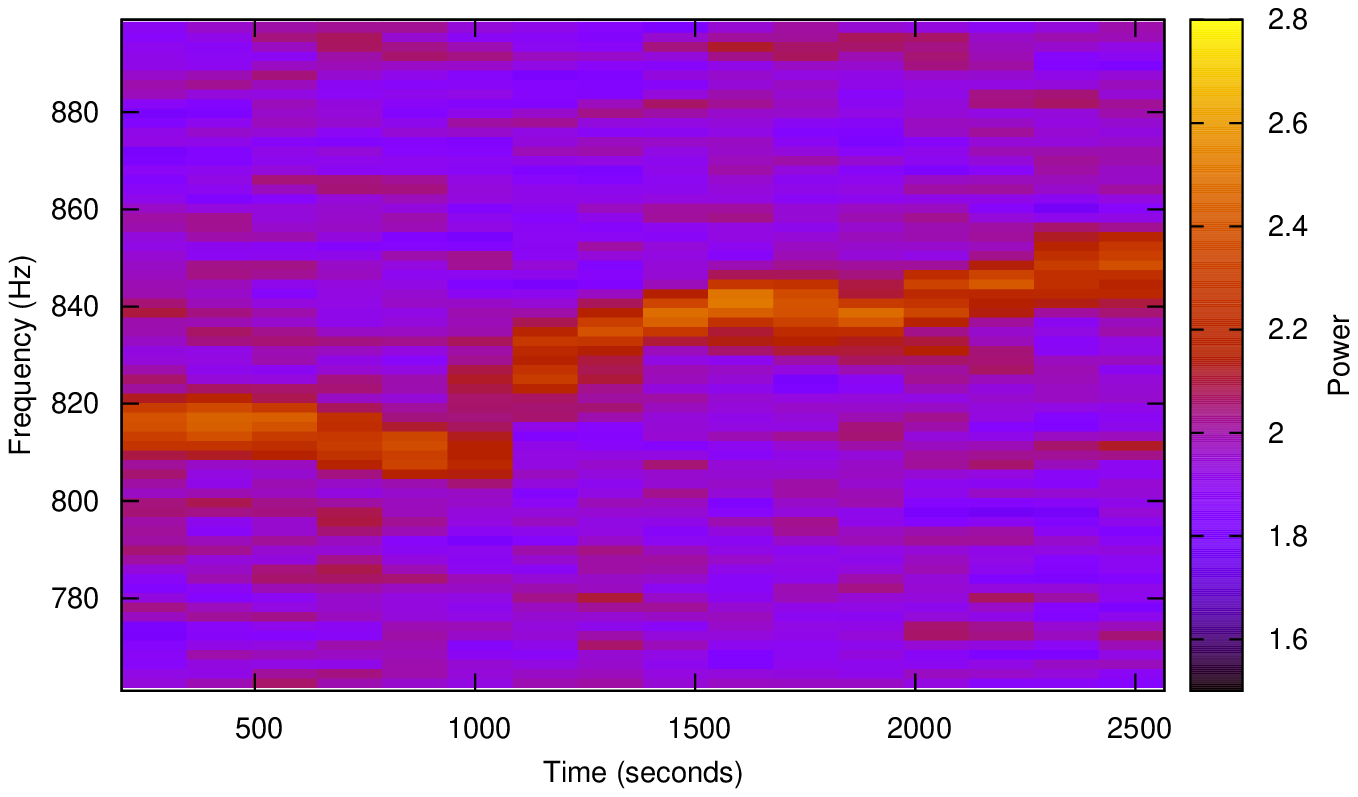} 
\includegraphics[width=0.3\textwidth,angle=-90]{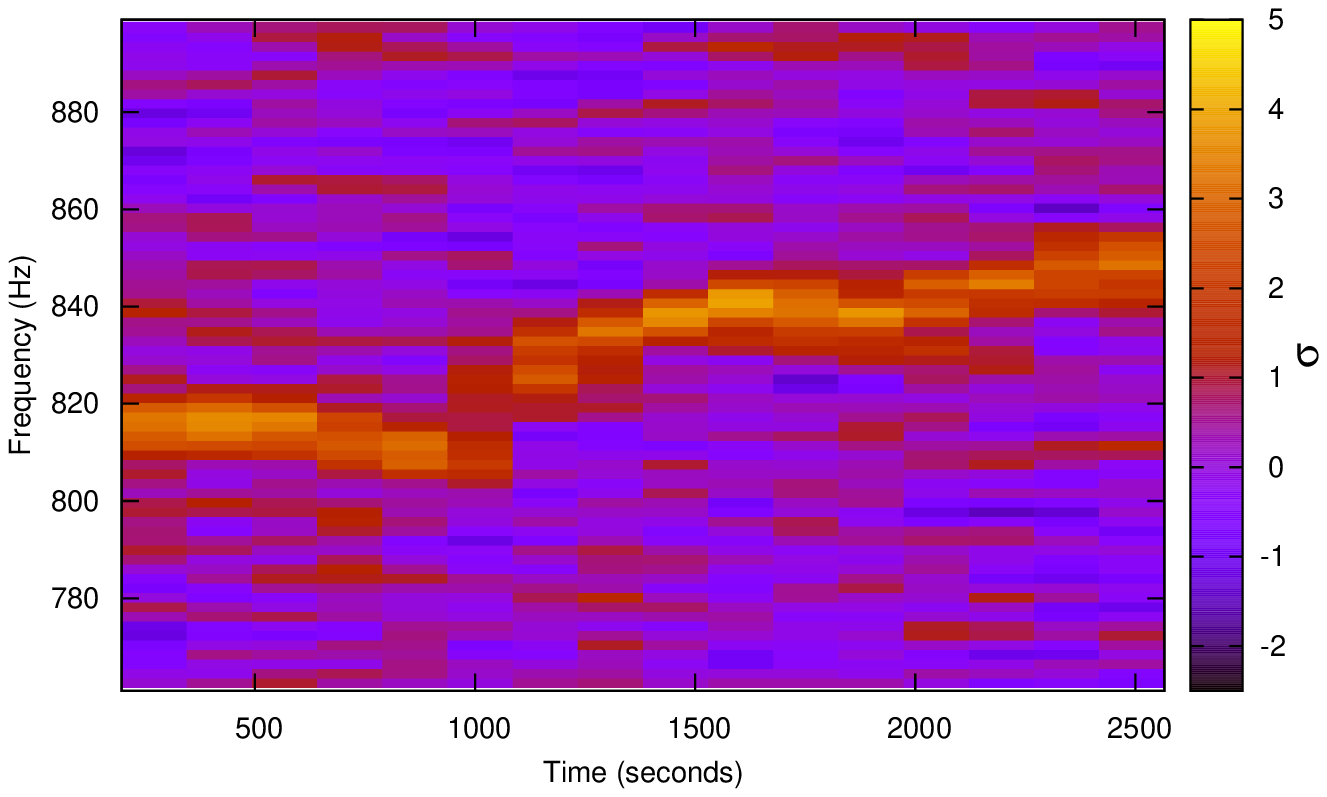}
\includegraphics[width=0.3\textwidth,angle=-90]{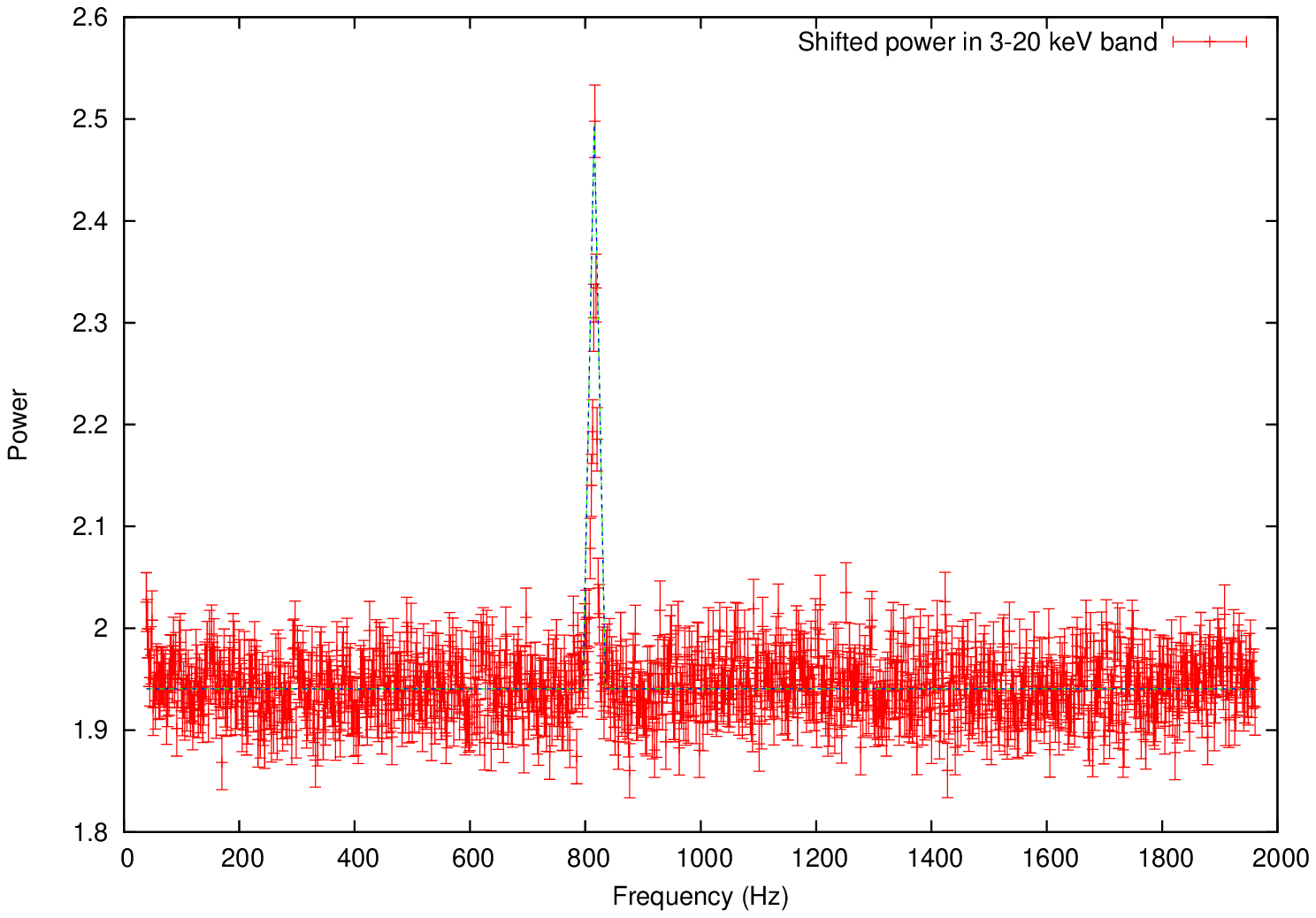}
\caption{Dynamic power spectra of 4U 1728-34 in the energy
range 3--20 keV (top panel). A drifting kHz QPO whose frequency changes
from $\sim 815$ Hz to $\sim 850$ Hz is clearly visible. The middle panel shows
the variation of $\sigma \equiv (P - P_n)/\Delta P$ showing that the QPO is
significantly detected. The bottom panel shows the co-added power spectra after
aligning the QPO frequency. No other features are detected in the 200 - 2000 Hz.  }
\label{Dyn_kHz}
\end{figure}

\begin{figure}
\centering
\includegraphics[width=0.3\textwidth,angle=-90]{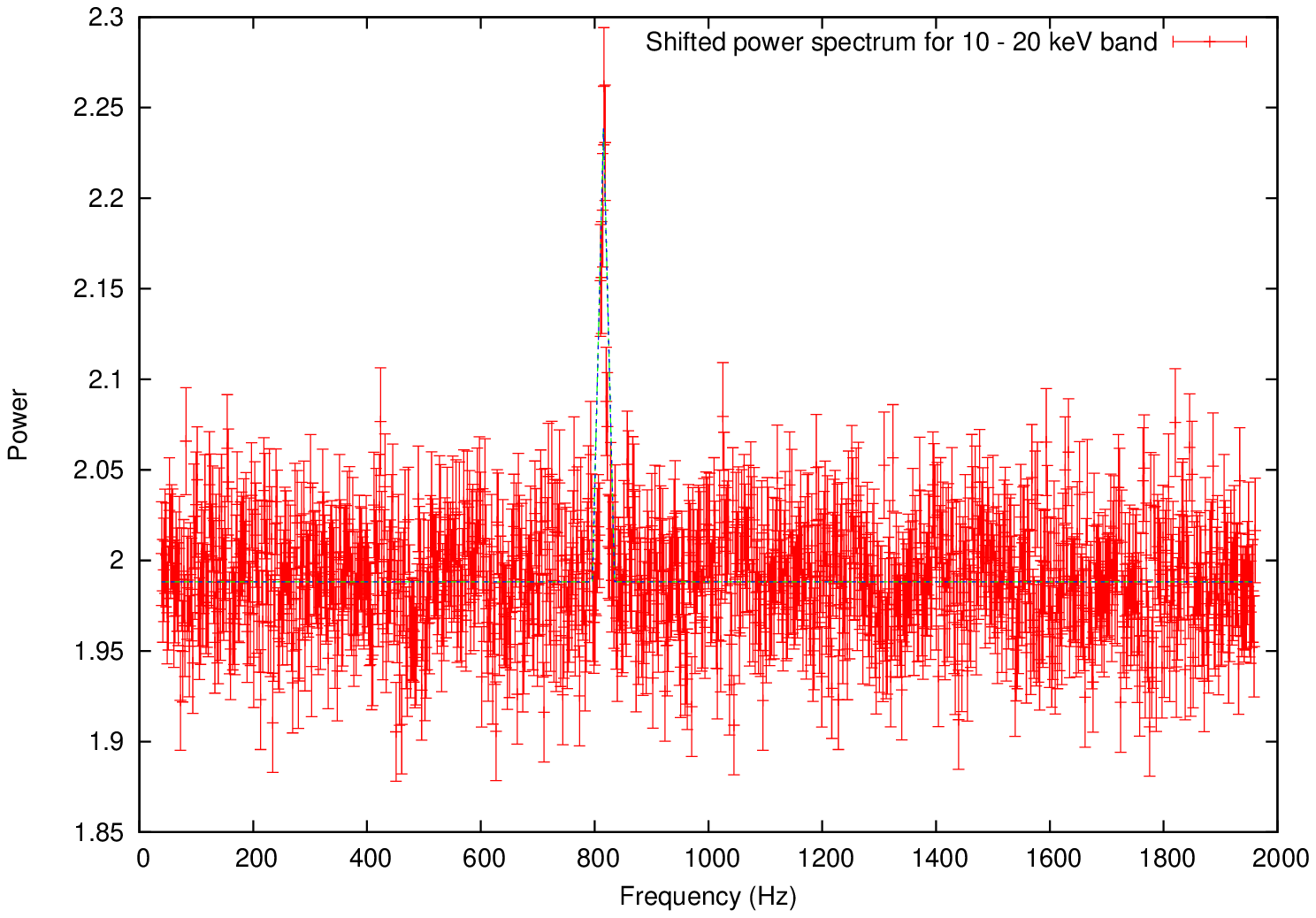}
\includegraphics[width=0.3\textwidth,angle=-90]{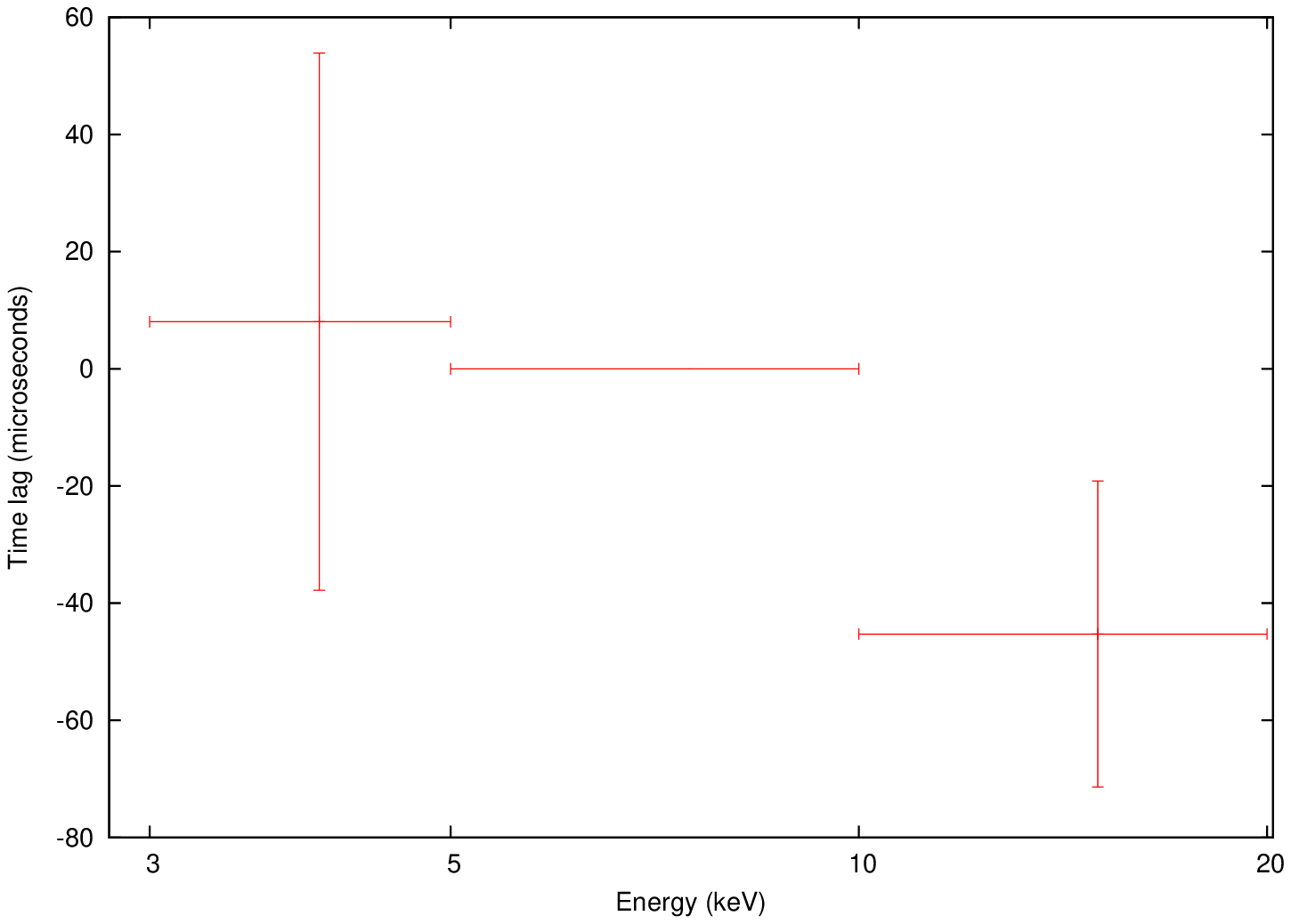}
\caption{Time lag as a function of energy for the kHz QPO. The
reference band here is  5-10 keV. Even for the short duration of
$\sim 2500$ seconds data the time-lag can be constrained to $< 100$
microseconds. }
\label{timelag}
\end{figure}

 The 1 second binned { {light curve}} generated using
data from {\it AstroSat} orbit number 2398 during March 8th  2016
is shown in Figure \ref{lightcurve}. 4U 1728-34
 was detected at a level of $\sim 1500$ c/s and near the end of the 
observation there was a Type 1 burst where the flux reached $\sim 10000$ 
c/s at the peak.

Figure \ref{spec} shows the photon spectrum from one of the LAXPC units
(namely LAXPC 10) for the first $\sim 2500$ secs of data. The response matrix
and the background were obtained using software which would be part of
the standard LAXPC pipeline and as described in \citet{Yad16,Ant16} and a systematic
uncertainty of 1\% was included in the spectral fitting. The data can
be reasonably modelled ($\chi^2$/dof = 92.3/104) by an 
absorbed thermal Comptonization component
(``Tbabs*nthcomp'' in Xspec) with photon index $1.8\pm0.02$ and temperature
 $3.05\pm0.04$ keV and a column density of $2.2\pm 0.4 \times 10^{22}$ cm$^{-2}$.
A weak but broad Iron line is required for the fit. Figure  \ref{spec} also shows the expected background spectrum and we note that it does not
dominate till $\sim 20$ keV.  Given the spectral
resolution and uncertainty in the response, the spectral shape of the
source is as expected and we concentrate now on the
rapid timing behaviour which is the focus of the present work.

The power spectrum for the first 2500 secs showed evidence for features
around $\sim 800$ Hz which suggested the presence of a drifting kHz QPO.
This was confirmed using dynamic power spectra analysis; the results of
which are  shown in top panel of Figure \ref{Dyn_kHz}. The dynamic power spectra were
created by splitting the  3-20 keV light curve into 16 parts of 147.97 seconds each.
Each part was then divided into 289 segments of 0.512 seconds. The power spectra
have been normalized such that the Poisson level, $P_N$ is at 2 i.e.  they are 
``Leahy'' normalized \citep{Lea83}.  The $\sim$40 microsecond dead-time of the
instrument reduces the noise level slightly to $\sim 1.95$. 
Power spectra were created for each of the 289 segments and averaged. Hence,
the error on the power at each frequency, $\Delta P/P$  
is $1/\sqrt(N) = 1/\sqrt(289)$ or 5.9 \%. The middle panel of  Figure \ref{Dyn_kHz} 
shows the significance $\sigma \equiv (P - P_N)/\Delta P$ for the detections.
The Figures clearly reveal a significant QPO whose frequency drifts from  
from $\sim 815$ Hz at the beginning of the observation to $\sim 850$ towards the end.
To explore the possibility of any other QPO in the data, we used the standard shift
and add technique where the power spectrum for each part is shifted 
such that the QPO frequencies becomes aligned and then averaged \citep{Men98b,Muk12}. The resultant power
spectrum is shown in the bottom panel of Figure \ref{Dyn_kHz} which shows no other
QPO like features.

We test whether  the kHz QPO is also detected at high energies ($> 10 keV$), especially
since earlier RXTE analysis of the source was unable to do so \citep{Muk12}. Power spectra were computed in the 10-20 keV band and following \citet{Muk12}, the spectra for each
part were shifted in frequency using the QPO detected in the 3-20 keV spectra as 
the reference. The avareged spectrum is shown in the top panel of Figure 
\ref{timelag} and the khz QPO is clearly detected in the high energy band. Fitting the
power spectrum with a Gaussian and a constant component 
gives $\chi^2$/dof $ = 991.2/982$ while only a constant component gives  $1231.8/984$ or
a $\Delta \chi^2 = 240$ for 2 additional degrees of freedom.

One does not expect to obtain tight constrains on energy dependent
time-lag from such a short duration data, especially when the frequency
of the QPO is drifting. Nevertheless, the detection of the QPO in high energy bands allows the
computation of energy dependent time-lags as shown in Figure \ref{timelag}. The time-lag
was computed from the shift and added averged cross-spectra as described for the power
spectra above and its error was estimated using the method described in \citet{Now99}.
Since the length of each segment is 0.512 secs, the frequency resultion $\Delta f$ of the
cross-spectra is $\sim 1.98$ Hz. We have verified that changing this resolution to $\sim 4$ or
$\sim 1$ Hz does not change the results obtained.
 The time-lag is constrained
to be less than 100 microseconds which indicates the capabilities of
the LAXPC to undertake such analysis with larger or better quality
observations.

\section{Detection of Burst Oscillations}

\begin{figure}
\centering
\includegraphics[width=0.34\textwidth,angle=-90]{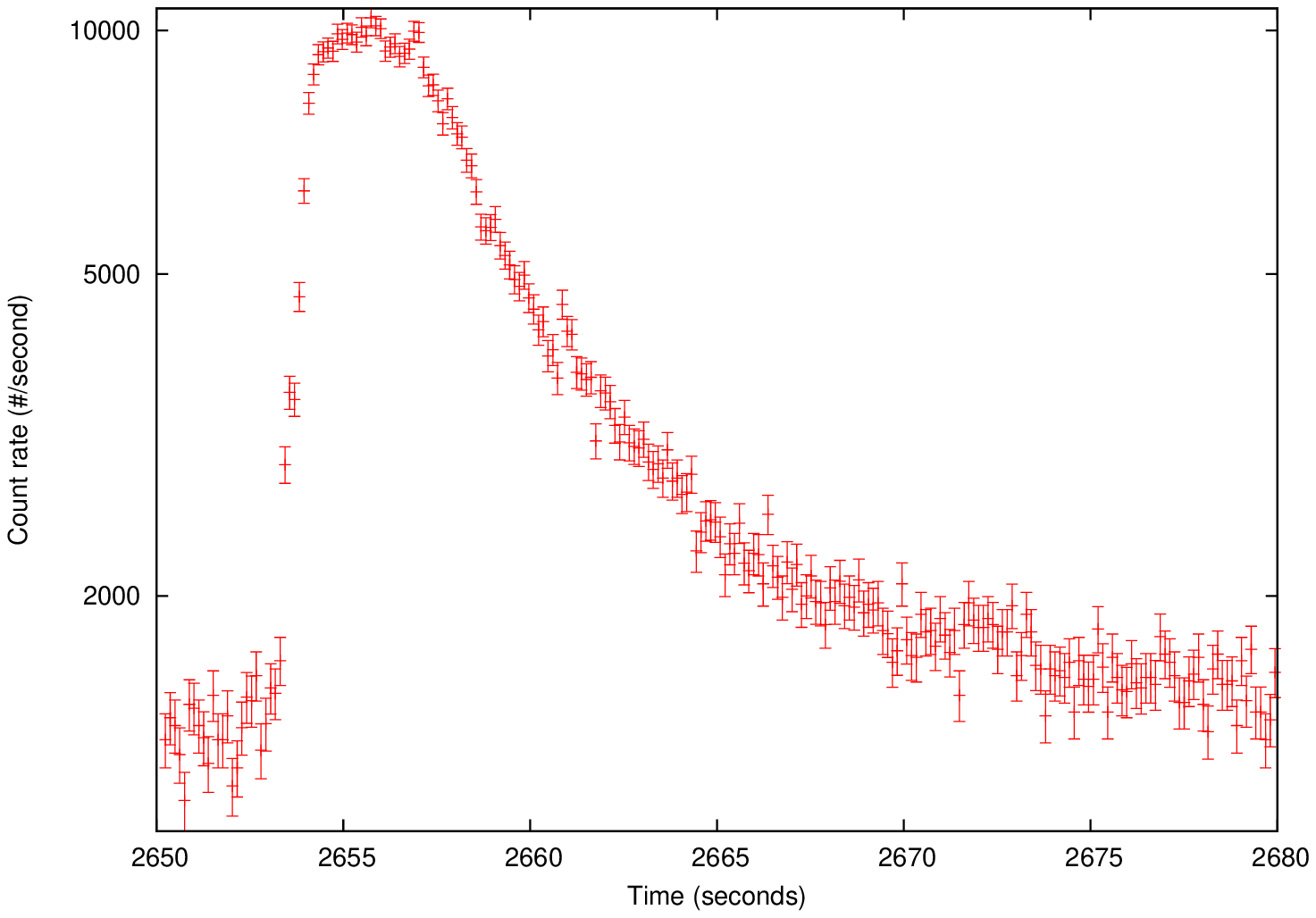}
\caption{Lightcurve of  the Type I X-ray burst observed in
4U 1728-34 in the energy
range 3--20 keV. The count rate from all three LAXPC
detectors are combined and the time bin is 0.128 seconds. }
\label{lightcurve_burst}
\end{figure}

\begin{figure}
\centering
\includegraphics[width=0.3\textwidth,angle=-90]{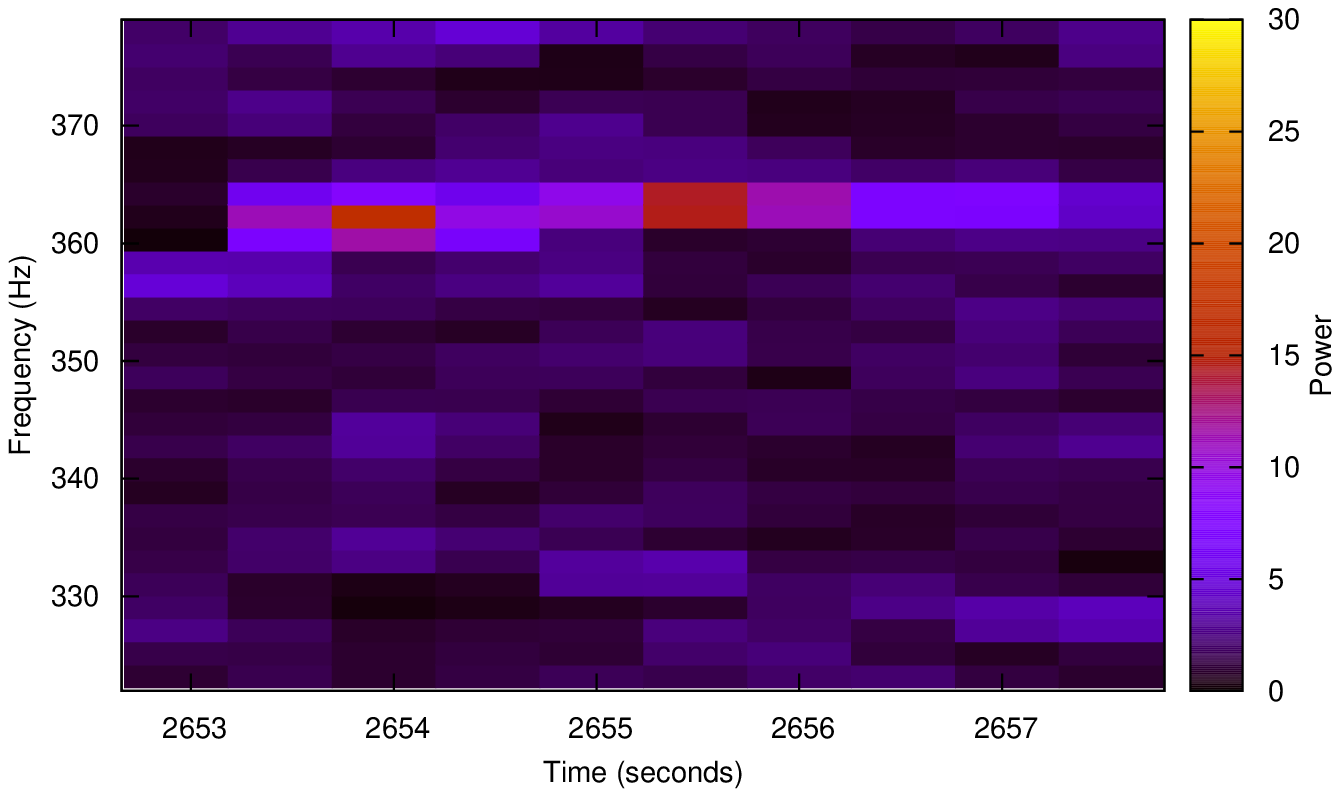}
\includegraphics[width=0.3\textwidth,angle=-90]{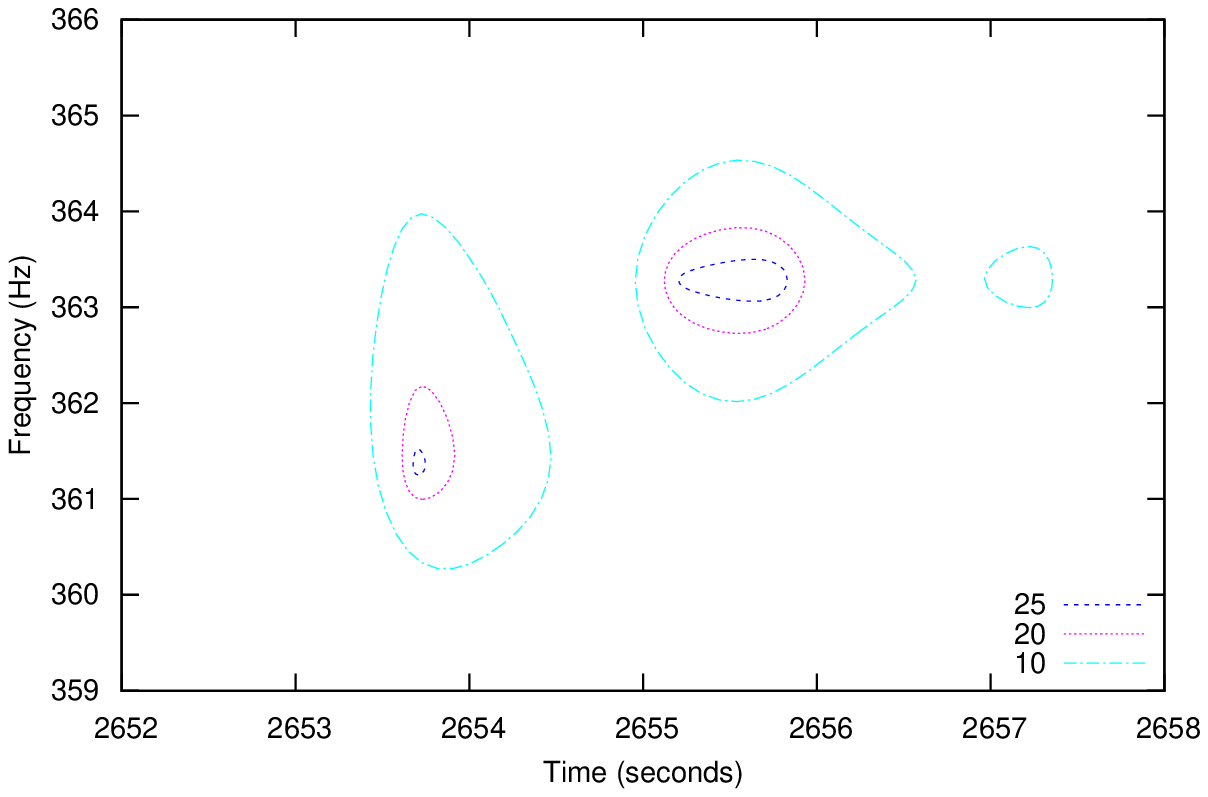}
\caption{Dynamic power spectra of the first 5 seconds of the Type 1 burst shown in color map
(top panel) and contour representation (bottom panel). {\color {blue} The power spectrum are ``Leahy'' normalized and
the contour lines are drawn for power values of 10, 20 and 25.}
A coherent feature at 363 Hz is seen.  }
\label{BO_dyn}
\end{figure}

Figure \ref{lightcurve_burst} shows the lightcurve of the Type-1
burst at a finer time resolution of 0.128 seconds. The burst profile
is typical with a fast rise and slower decay lasting for about
20 seconds. Since burst oscillations are often detected in the
early phase of the burst, we looked for high frequency signatures
in the first 5 seconds of the burst by computing the dynamic
power spectra.  The first 5.12 seconds of the burst was divided into 
10 parts and the power spectra were computed for each of them to get the
dynamic power spectra as shown in the top panel of Figure \ref{BO_dyn}.
A coherent feature is observed at
362 Hz which is referred to as the Burst Oscillation.  The power spectra are ``Leahy'' normalized and each one is computed from one segment. Thus the null hypothesis probability that a power as high as observed is obtained by chance
is $N_f e^{(-P(f)/2)}$ \citep[e.g.][]{Vau94}, where $N_f$ is the number of independent frequencies being considered. Since we were searching for oscillations $< 1000$ Hz at
a frequency resolution of $\sim 2$ Hz, we chose $N_f = 500$. The maximum value of $P(f)$ is $\sim 34$,
which implies a null hypothesis probability of $2 \times 10^{-5}$ . Thus the oscillation is detected at a high level
of significance. The bottom panel of Figure \ref{BO_dyn} shows a more close up view of the Dynamic power spectra represented by a contour map. The Burst Oscillation frequencies increases from $\sim 361.5$ to $\sim 363.5$ Hz which has been reported earlier using RXTE for this
and other sources 
\citep[e.g.][and references therein]{Wat12}.

\section{Discussion} 

We have presented here the first detection of both kinds of milli-second variability i.e kHz QPOs and Burst Oscillation in the LMXB 4U 1728-34 with {\it AstroSat}/LAXPC from a single $\sim 3$ ksec observation. This result demonstrates that the LAXPC instrument has the necessary sensitivity and time resolution to detect and study millisecond timing phenomenon in 3--20 keV and possibly at higher energy also. 

With RXTE, there have been relatively few studies 
of the high frequency QPOs at energy $\>$20 keV mainly due 
to rapid decline in the
        effective of PCA at higher energies. In fact, detailed energy dependence of both the fractional r.m.s
and time lag for kHz QPO have been possible for only one or two orbits of RXTE observations i.e. the March 3rd 1996 observation
of 4U 1608-52 \citep{Ber96,Vau97} and February 24 1998 and  April 27 1996 observations of 4U 1636-53 \citep{Zha96,Kaa99}. \citet{Bar13}, \citet{Ave13} and \cite{Pei15} had to combine a large number of RXTE observations to obtain average energy dependent r.m.s and time-lag for different QPO frequency ranges. 
During its latter stage, RXTE observations were undertaken with 
only one or two of its five PCUs, hence making significant detections in short observations harder. For example, for the source analyzed in this work,
4U 1728-34, the QPO was not detected in the 10-20 keV band even when {$~$85 ksec} were analyzed \citep{Muk12}. In contrast, the 3 ksec LAXPC data detected the QPO in the band as shown in the dynamic power spectrum (Figure \ref{Dyn_kHz}). Thus  {\it AstroSat}/LAXPC has now proven potential to study the energy dependence of kHz QPO.
A critical advantage will be obtained by the {\it  simultaneous} observations from other instruments on board
AstroSat, especially the Soft X-ray Telescope (SXT). The broad band coverage will measure more accurately
the time-averaged radiative components of the system thereby having significantly more constrains. For example,
using RXTE observations the spectral modelling was degenerate leading to ambiguities regarding the size and
geometry of the source as inferred from the energy dependent properties of the kHz QPO \citep{Kum16}. The SXT
spectra in the 0.3--8 keV band will lift this degeneracy allowing one to test different models and to obtain
interesting physical constraints such as size and geometry of the source. 

Confirmation of the 363 Hz  burst oscillation of the source brings out the potential of LAXPC to
detect the phenomena and in general to enhance our understanding of Type-1 bursts.  AstroSat will also
enable the broad band study of the spectral evolution of the burst using LAXPC and SXT. Of particular
interest could be the measurement of time delays between the X-rays and the UV emission as detected
by the  Ultra-Violet Imaging Telescope (UVIT). These delays correspond to the light crossing
time to the outer disk and provide constrains on the disk geometry \citep[e.g.][]{Hye06}.

We defer detailed analysis and interpretation of the data to later works where the properties of the kHz QPO and Burst Oscillations will be studied separately.  Our first look results show with confidence that the {\it RXTE} legacy of studying milli-second variability of X-ray binaries will be 
effectively carried forward by {\it AstroSat} and one can look forward to new exciting discoveries in the near future.

\section{Acknowledgments}We acknowledge the strong support from Indian Space Research
Organization (ISRO) in various aspect of instrument building, testing,
software development and mission operation during payload verification
phase. { {We acknowledge support of the scientific and technical staff of the LAXPC team as well as staff of the TIFR Workshop in the development and testing of the LAXPC instrument.}}

\end{document}